\documentclass[letterpaper]{article}

\usepackage{aaai}
\nocopyright

\usepackage{times}
\usepackage{helvet}
\usepackage{courier}

\usepackage{graphicx}
\usepackage{epsfig}
\usepackage{amssymb}
\usepackage{epsfig}
\usepackage{parskip}
\usepackage{ntheorem}
\usepackage{amsmath}
\usepackage{verbatim}
\usepackage{centernot}

\usepackage{subfig}

\usepackage{syntax}

\newcommand{\ci}{\perp\!\!\!\perp}
\newcommand{\nci}{\centernot{\ci}}

\begin{document}

\title{Whittemore: An embedded domain specific language for causal programming}
\author{Joshua Brul\'{e}\\
Department of Computer Science\\ University of Maryland, College Park\\
jbrule@cs.umd.edu}

\maketitle
\begin{abstract}
\begin{quote}
    This paper introduces Whittemore, a language for causal programming. Causal programming is based on the theory of structural causal models and consists of two primary operations: \emph{identification}, which finds formulas that compute causal queries, and \emph{estimation}, which applies formulas to transform probability distributions to other probability distribution. Causal programming provides abstractions to declare models, queries, and distributions with syntax similar to standard mathematical notation, and conducts rigorous causal inference, without requiring detailed knowledge of the underlying algorithms. Examples of causal inference with real data are provided, along with discussion of the implementation and possibilities for future extension.
\end{quote}
\end{abstract}

\section{Introduction}

In computer science, programming has come to refer to two related concepts. Programmers generally think of programming as describing a process to be executed, usually by writing a sequence of instructions, i.e. code. Mathematicians often think of programming in the sense of linear programming: declaring a problem in terms of satisfiability and/or optimization criteria, to be solved by particular algorithms.

The `mathematical' and `code' senses of programming overlap in practice. Some notable families of programming languages that realize the more mathematical aspect of programming include logic/relational programming \cite{byrd2009}, and probabilistic programming \cite{mansinghka2009}. Logic programming is largely based on defining formulas in first-order logic. Probabilistic programming can be seen a language for defining probability distributions, with an operator that implements conditional sampling \cite{goodman2008}.

This paper introduces ``causal programming'', based on the theory of Pearlian structural causal models (SCM) and related inference rules and algorithms \cite{pearl1995} \cite{shpitser2008}. The key theoretical contribution is to define causal programming as an abstraction over two primary operations: identification, which finds formulas that compute a causal query of interest, and estimation, which applies formulas to transform probability distributions to other probability distributions.

This paper describes the causal programming language ``Whittemore'', which is implemented as an embedded domain specific language. The syntax and semantics of Whittemore are designed represent the underlying mathematical concepts as closely as possible. Readers familiar with both structural causal models and Lisp may wish to go directly to the ``Examples and interactive computing" section.



\section{Syntax and semantics}

Whittemore is defined as a total (i.e. always terminating), purely functional subset of its host language. The reference implementation of Whittemore is in Clojure, a dialect of Lisp \cite{hickey2008}. This paper describes Whittemore using the same notation for expressions and data types as Clojure and generally follows the same conventions as idiomatic Clojure code.

An expression in Whittemore is a constant, symbol, or \mbox{\texttt{(\emph{op} \emph{expr}*)}} where \emph{op} is a causal programming operator, and \emph{expr} is an expression. Operators are described using regular expression syntax: \texttt{?} (optional), \texttt{*} (0 or more), \texttt{+} (1 or more), with non-terminals denoted by \emph{italics}.

\begin{figure}[h]
\begin{grammar}
    <expr> ::= <constant> | <symbol> | (<op> <expr>$^*$)

    <op> ::= define | model | data | q | identify
    \alt estimate | measure | signature | <distribution>
\end{grammar}
    \caption{Whittemore grammar}
\end{figure}

\subsection{Constants}

Constants include standard atomic data types (e.g. integer and floating point numbers, strings, booleans) as well as keywords, which are symbolic identifiers that evaluate to themselves. Keywords begin with a colon and can contain alphanumeric characters and special characters that are not reserved by the host language, e.g. \texttt{:x, :x', :treatment, :z\_1} are all valid keywords.

In addition to the atomic data types, constants include the following collection types, with literal syntax:

\begin{itemize}

\item
Vectors are ordered collections of values, e.g. \texttt{[:x :y]}

\item
Maps are unordered collections that maps unique keys to values, e.g. \texttt{\{:x 0, :y 1\}}

\item
    Sets are unordered collection of unique values, e.g. \mbox{\texttt{\#\{:x :y\}}}.

\end{itemize}

Keywords and sets are optional data types, in that strings can generally be used in place of keywords, and vectors can be used in place of sets, without affecting the semantics of the program. However, keywords and set notation are preferred in some cases where it is useful to have a visual distinction.

\subsection{Symbols}

\texttt{(define \emph{symbol} \emph{docstring?} \emph{value})}

Symbols are identifiers that normally refer to another value. The \texttt{define} operator binds a symbol to a value, and returns the value. Note that \texttt{define} cannot rebind symbols, which is necessary for Whittemore to be a purely functional language.

\subsection{Identification operators}

The identification operators correspond to the task of identification from population distributions, i.e. the limit of infinite samples \cite{heckman2005}.

\subsubsection{Model}
\texttt{(model \emph{dag} \emph{confounding}*)}

A Model corresponds to the concept of a semi-Markovian causal diagram \cite{shpitser2008}, representing a class of structural causal models. The \texttt{model} operator returns a new Model where \emph{dag} is a map of variables to their parents, and \emph{confounding} is a set of endogenous variables whose background variables are not independent.

\begin{figure}[h]
    \begin{minipage}{.5\linewidth}
\centering
\begin{verbatim}
(define front-door
  (model
    {:x []
     :z [:x]
     :y [:z]}
    #{:x :y}))
\end{verbatim}
\end{minipage}%
    \begin{minipage}{.5\linewidth}
\centering
\begin{align*}
    X &= f_X(\epsilon_X) \\
    Z &= f_Z(x, \epsilon_Z) \\
    Y &= f_Y(z, \epsilon_Y) \\
    \epsilon_X &\nci \epsilon_Y
\end{align*}
\end{minipage}
\begin{minipage}{\linewidth}
\includegraphics{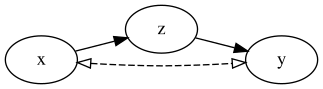}
\end{minipage}

    \caption{An expression defining a model, equivalent mathematical notation, and the resulting causal diagram.}
\end{figure}

\subsubsection{Data}
\texttt{(data \emph{joint})}

Data represents the `signature' of a probability function, i.e. symbolic information about a population probability distribution that is required by the underlying causal inference algorithms. Whittemore currently only supports representing knowledge of joint probability functions, for example, knowledge of the joint probability function, $P(x, y, z)$, is represented as \mbox{\texttt{(data [:x :y :z])}}.

\subsubsection{Query}

\texttt{(q \emph{effect} :do \emph{do}? :given \emph{given}?)}

A Query is a statistical or causal query, such that the resulting value is a probability distribution. For example, \mbox{\texttt{(q [:y] :given \{:x 1\})}} corresponds to \mbox{$P(y \mid X=1)$}, a statistical query. \mbox{\texttt{(q [:y\_1 :y\_2] :do \{:x 0\})}} corresponds to $P(y_1, y_2 \mid do(X=0))$, a causal query. Whittemore does not currently support counterfactual queries, although support is planned for a future release.

Note that since \emph{do} and \emph{given} are both optional, they are implemented as keyword arguments in the host language. Their default values are the empty map.

\subsubsection{Formula} \texttt{(identify \emph{model} \emph{data}? \emph{query})}

The \texttt{identify} operator returns a Formula that computes \emph{query}, as a function of \emph{data}, in every SCM entailed by \emph{model}, or a Fail, if such a Formula does not exist. If unspecified, \emph{data} defaults to the joint observational probability function over all endogenous variables in \emph{model}. For example, \mbox{\texttt{(identify front-door (q [:y] :do \{:x 0\}))}}, with implicit \texttt{(data [:x :y :z])}, returns the Formula:
\begin{align*}
    &\sum_{z} \left[ \sum_{x} P(y \mid x, z) P(x) \right] P(z \mid x)\\
    &\text{where: } x=0
\end{align*}

Note that Formulas follow lexical scoping rules, e.g. only the `outer' $x$ is bound to $0$. The implementation of Formulas is discussed in the ``Implementation'' section.

The same \texttt{identify} expression, but with \mbox{\texttt{(data [:x :y])}} would return a Fail describing the hedge \cite{shpitser2008} that renders identification impossible. Since \texttt{identify} is based on the ID algorithm \cite{shpitser2006-identification}, it is complete; a Fail will be returned if and only if no appropriate Formula exists.

\subsection{Distributions}

The causal programming concept of a distribution corresponds to the mathematical concept of a probability distribution. However, an `impedance mismatch' between formulas and distributions exists. Formulas represent the transformation of probability distributions to probability distributions --- usually, an observational to an interventional distribution. However, the structural causal model approach to causal inference is entirely nonparametric, whereas the evaluation of expressions of random variables requires specific knowledge of their underlying distributions.



The core problem is to provide an abstraction over the general concept of a probability distribution, while implementing distribution-specific computations. Whittemore's solution is to rely on dynamic polymorphism, dispatching on the type of the distribution. A Distribution implementation respects the following protocol:

\begin{itemize}
    \item
        \texttt{(\emph{distribution} \emph{expr}*)}\\ Returns an instance of the probability distribution.

    \item \texttt{(estimate \emph{this} \emph{formula})}\\ Returns the result of applying a \emph{formula} to \emph{this} distribution, yielding a new distribution. Note that a Query acts as a special case of a Formula.

    \item \texttt{(measure \emph{this} \emph{event})}\\
        Returns the probability of \emph{event}, i.e. \texttt{measure} implements the mathematical concept of a probability measure. An \emph{event} is expected to be a map of keywords to values.

    \item \texttt{(signature \emph{this})}\\
        Returns the Data `signature' of the distribution.

\end{itemize}



Whittemore includes an implementation of a categorical distribution which accepts a single argument of a vector of samples (events) and infers the support of the support of the distribution. For example:

\begin{verbatim}
(define example-distribution
  (categorical
    [{:x 0, :y 0}
     {:x 0, :y 1}
     {:x 1, :y 0}
     {:x 1, :y 1}
     {:x 1, :y 1}]))
\end{verbatim}

Binds a representation of a probability distribution where $P(x, y) = 2/5$ and \mbox{$P(x', y)=P(x,y')=P(x', y')=1/5$} to the symbol \texttt{example-distribution}.

The Distribution protocol is user extensible; other probability distributions can be implemented in the host language without modification to Whittemore's implementation.

\subsection{Infer and `syntactic sugar'}

The reference implementation of Whittemore provides some `syntactic sugar' to make causal programming easier. In particular, the \texttt{q} operator has three versions that mimic common usage of $P()$ in probability theory:

\begin{itemize}
    \item `Unbound' query, e.g. \texttt{(q [:y] :do [:x])}, a query where \emph{do} and \emph{given} are vectors. An unbound query can still be provided as an argument to \texttt{identify}, but the resulting formula cannot be used as an argument to \texttt{estimate} without first providing the necessary variable bindings.
    \item `Bound' query, e.g. \texttt{(q [:y] :do \{:x 0\})}, corresponding to a conditional or interventional distribution (this is considered the canonical version of a Query).
    \item `Event' query, e.g. \texttt{(q \{:y 1\} :do \{:x 0\})}, corresponding to a specific probability, i.e. \emph{effect} is an event.
\end{itemize}

Providing an event query to \texttt{estimate} implies \texttt{measure}. For example, assuming that an appropriate probability distribution is bound to the symbol \texttt{smoking}\footnote{These examples assume that \texttt{smoking} follows the probability distribution in \cite[Table~3.1]{pearl2009}}:

\begin{verbatim}
(estimate smoking
  (q {:y 1} :given {:x 1}))
\end{verbatim}
Returns the probability $0.8525$.

In addition, Whittemore provides the \texttt{infer} operator, which combines the functionality of \texttt{identify}, \texttt{estimate} and \texttt{measure}. For example:

\begin{verbatim}
(infer front-door smoking
  (q {:y 1} :do {:x 1}))
\end{verbatim}
Returns the probability $0.4975$.

\section{Examples and interactive computing}

Whittemore is designed to be used interactively in a notebook interface. The reference implementation of Whittemore has built-in support for Jupyter \cite{jupyter}, an open-source, interactive computational environment. Jupyter integration supports rich output; models are automatically displayed as causal diagrams, and formulas are displayed as \LaTeX math. The code examples in this section are shown along with their Jupyter notebook outputs.

These examples use some additional functions that are not part of `core' Whittemore. To load data, \texttt{read-csv} parses and processes a comma-separated values file; \texttt{head} returns the first $n$ samples for inspection. To visualize a probability distribution, \texttt{plot-univariate} returns a plot of a marginal distribution.

\begin{figure}[h]
\begin{verbatim}
(define kidney-dataset
  (read-csv "data/renal-calculi.csv"))

(head kidney-dataset 5)

\end{verbatim}
\includegraphics[width=0.6\linewidth]{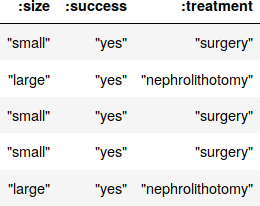}
    \caption{The first 5 rows of ``renal-calculi.csv''}
\end{figure}

\begin{figure}[h]
\begin{verbatim}
(define kidney-distribution
  (categorical kidney-dataset))

(plot-univariate
  kidney-distribution :success)

\end{verbatim}
\includegraphics[width=0.8\linewidth]{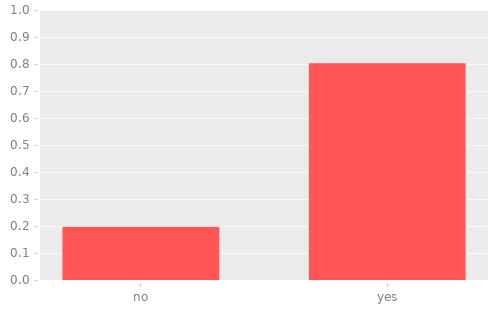}
    \caption{A plot of the marginal distribution of \texttt{:success}}
\end{figure}

\subsection{Resolving Simpson's paradox}

The \texttt{kidney-distribution} is the empirical probability distribution associated with a study of treatment of renal calculi, i.e. kidney stones \cite{charig1986}. There are three categorical random variables: $\text{treatment} \in \{\text{``surgery"}, \text{``nephrolithotomy"}\}$, $\text{size} \in \{ \text{``small''}, \text{``large''} \}$, and $\text{success} \in \{ \text{``no''}, \text{``yes''} \}$. This distribution exhibits Simpson's paradox, which despite being well known, still continues to ``trap the unwary'' \cite{dawid1979} \cite{pearl2014}.

\begin{minipage}{\linewidth}
In this distribution, the probability of success, given surgery, is less than the probability of success, given nephrolithotomy:\\

\begin{verbatim}
(estimate kidney-distribution
  (q {:success "yes"}
     :given {:treatment "surgery"}))
\end{verbatim}
0.78\\
\begin{verbatim}
(estimate kidney-distribution
  (q {:success "yes"}
    :given {:treatment "nephrolithotomy"}))
\end{verbatim}
0.8257142857142857
\end{minipage}

However, a reversal appears when conditioning on subgroups. When restricted to observing patients with small kidney stones, surgery appears to be the superior treatment:

\begin{minipage}{\linewidth}
\begin{verbatim}
(estimate kidney-distribution
  (q {:success "yes"}
     :given {:size "small"
             :treatment "surgery"}))
\end{verbatim}
0.9310344827586207\\

\begin{verbatim}
(estimate kidney-distribution
  (q {:success "yes"}
    :given {:size "small"
            :treatment "nephrolithotomy"}))
\end{verbatim}
0.8666666666666667
\end{minipage}

And when restricted to observing patients with large kidney stones, surgery again appears to be the superior treatment:

\begin{minipage}{\linewidth}
\begin{verbatim}
(estimate kidney-distribution
  (q {:success "yes"}
    :given {:size "small"
            :treatment "surgery"}))
\end{verbatim}

0.7300380228136882\\

\begin{verbatim}
(estimate kidney-distribution
  (q {:success "yes"}
    :given {:size "small"
            :treatment "nephrolithotomy"}))
\end{verbatim}

0.6875
\end{minipage}

Researchers familiar with the do-calculus will immediately recognize that resolving the question of which treatment is superior is a causal, not statistical query. The aim of causal programming is to answer such queries without the need for the user to understand the details of causal inference. The only requirement is to specify the causal diagram corresponding to the model that produced the original distribution.

\begin{figure}[h]
\begin{verbatim}
(define charig1986
  (model
    {:size []
     :treatment [:size]
     :success [:treatment :size]}))

\end{verbatim}
\includegraphics[width=\linewidth]{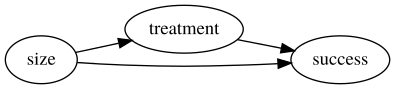}
    \caption{A model where kidney stone size affects the success of treatment and which treatment a patient receives}
\end{figure}

\begin{minipage}{\linewidth}
With respect to the casual assumptions in \texttt{charig1986} \cite{charig1986}, an estimate of the casual effect of treatment on success is easily inferred:\\

\begin{verbatim}
(infer charig1986 kidney-distribution
  (q {:success "yes"}
     :do {:treatment "surgery"}))
\end{verbatim}
0.8325462173856037\\

\begin{verbatim}
(infer charig1986 kidney-distribution
  (q {:success "yes"}
     :do {:treatment "nephrolithotomy"}))
\end{verbatim}
0.778875
\end{minipage}

\subsection{Nonstandard adjustments}

Note that Whittemore is by no means limited to the special cases of back door and front door adjustment \cite{pearl1995}. Causal programming easily identifies formulas for computing causal effect that involve non-standard adjustments:

\begin{figure}[h]
\begin{verbatim}
(define concomitant-example
  "Figure 1 (f) from (Shpitser 2008)"
  (model
    {:y [:x :z_1 :z_2]
     :z_2 [:z_1]
     :z_1 [:x]
     :x []}
    #{:y :z_1}
    #{:x :z_2}))

\end{verbatim}
\includegraphics[width=\linewidth]{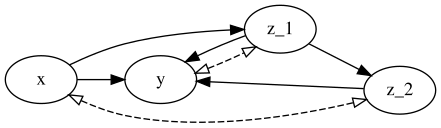}
    \caption{A model where the identification of $P(y \mid do(x))$ requires summing/integrating over $z_1$ and $z_2$}
\end{figure}

\begin{verbatim}
(identify concomitant-example
  (q [:y] :do [:x]))
\end{verbatim}
\[
\sum_{z_1, z_2} \left[ \sum_{x} P(x) P(z_2 \mid x, z_1) \right] P(z_1 \mid x) P(y \mid x, z_1, z_2)
\]

\section{Implementation}

The ID algorithm and several related algorithms have been previously implemented in the R programming language \cite{tikka2017}. Whittemore's \texttt{identify} is a purely functional implementation of ID, designed to be easily extensible.

The Model, Data, Query and Formula types are all implemented as persistent (immutable) hash array mapped tries \cite{bagwell2001} which support lookup and `modification' (associating a key and value creates a new data structure) in $\log_{32} N$ time. This provides good performance while remaining free of side effects. A considerable advantage is that the data structures can be freely shared with any other part of a program --- it is impossible to corrupt a data structure since none of them can be changed.

Formulas are defined as a map of bindings of variables to values, and a \emph{form}, which is defined recursively:

\begin{itemize}
    \item \texttt{\{:p \#\{\emph{vars}\} :given \#\{\emph{vars}\}\}}
    \item \texttt{\{:sum \emph{form} :sub \#\{\emph{vars}\}\}}
    \item \texttt{\{:prod  \#\{\emph{forms}\}\}}
    \item \texttt{\{:numer \emph{form} :denom \emph{form}\}}
\end{itemize}

These forms correspond to a probability expression, summation, product, and fraction, respectively. Formulas follow lexical scoping rules, which obviates the need to rename variables --- variable bindings are determined the first surrounding \texttt{:sum} that contains the variable as a subscript.

Formulas can be manipulated with standard Clojure functions and are designed to support simplification in a manner similar to a nanopass compiler \cite{keep2012} where each `pass' takes a valid form and applies a simple rule to reduce the form to a simpler, still valid form. For example,

\begin{minipage}{\linewidth}
\begin{verbatim}
{:numer {:p #{:y :z :x}},
 :denom {:sub #{:y},
         :sum {:p #{:y :z :x}}}}
\end{verbatim}
\end{minipage}

Can be reduced to \texttt{\{:p \#\{:y\} :given \#\{:x :z\}\}} by applying a marginalization rule on the \texttt{:denom} form and then a conditioning rule on the resulting expression.

Additional keys can be added to the Model, Data, Query, and Formula types, without changing the semantics of a program, permitting considerable future extensibility.

\section{Discussion}

Whittemore demonstrates that the full causal inference `pipeline' --- from raw data to estimates of causal effect --- can be effectively abstracted over. Using Clojure as host language blurs the line between a programming language and library: Whittemore syntax is similar to standard mathematical notation, but remains a subset of legal Clojure expressions.

Whittemore currently only supports causal effect identification from observational probability distributions. There are several opportunities for future extension to `core' Whittemore:

\begin{itemize}

    \item Counterfactuals. A counterfactual query operator \texttt{(cf \emph{gamma} \emph{delta}?)} can be introduced without affecting the existing syntax and semantics. For example, \mbox{$P(Y_x \mid x')$}, the effect of treatment on the treated \cite{shpitser2009}, can be represented as \mbox{\texttt{(cf [:y \{:x 0\}] [\{:x 1\} \{\}])}}. Supporting counterfactual queries would require extending \texttt{identify} to implement the IDC* algorithm \cite{shpitser2007-counterfactuals}.

    \item Data fusion. ``Data fusion'' refers to the problem of causal inference given heterogeneous sources of data \cite{bareinboim2016-fusion}. This includes the problems of identification from surrogate experiments \cite{bareinboim2012-surrogate}, recovery from selection bias \cite{bareinboim2012-controlling}, and causal transportability \cite{bareinboim2013-transportability}. These problems could be represented as causal programs by adding additional Data signatures. For example, the availability of limited experimental data could be represented by \mbox{\texttt{(data \emph{joint} :do \emph{surrogate}?)}} and supported by extending \texttt{identify} to implement the zID algorithm \cite{bareinboim2012-surrogate}.
        
    \item Causal discovery. Causal discovery algorithms generally rely on information about conditional independences between variables \cite{malinsky2017}. This can be introduced to causal programming by extending the Data type to include such information.

\end{itemize}

The Distribution protocol is open to extension, without modifying the implementation of Whittemore. This provides an interesting opportunity to integrate casual programming with probabilistic programming. Probabilistic programming's primary operator is to efficiently sample from marginal and conditional probability distributions. Causal programming's primary operator is to identify causal effects in terms of such probabilities.

Whittemore is under active development. The reference implementation is cross-platform and open source: https://github.com/jtcbrule/whittemore

\bibliography{cpl}
\bibliographystyle{aaai}

\end{document}